\documentclass[prl,twocolumn,superscriptaddress,showpacs,floatfix,amsfonts]{revtex4}
\usepackage{graphicx,graphics,color,epsfig}
\usepackage{bm}
\usepackage{amsmath}
\usepackage{amssymb}
\begin{document}
\preprint{}
\title{Continuous Quantum Phase Transition in a Kondo Lattice Model}
\author{Jian-Xin Zhu}
\affiliation{Theoretical Division, Los Alamos National Laboratory,
Los Alamos, New Mexico 87545, USA}
\author{D. R. Grempel}
\affiliation{ CEA-Saclay/DRECAM/SPCSI, 91191 Gif-sur-Yvette,
France }
\author{Qimiao Si}
\affiliation{Department of Physics \& Astronomy, Rice University,
Houston, TX 77005--1892, USA }

\begin{abstract}
We study the magnetic quantum phase transition in an anisotropic
Kondo lattice model.
The
dynamical competition between the RKKY and Kondo interactions is
treated using an extended dynamic mean field theory (EDMFT)
appropriate for both the antiferromagnetic and paramagnetic
phases. A quantum Monte Carlo approach is used, which is able to
reach very low temperatures, of the order of 1\% of the bare Kondo
scale. We find that the finite-temperature magnetic transition,
which occurs for sufficiently large RKKY interactions, is first
order. The extrapolated zero-temperature magnetic transition, on
the other hand, is continuous and locally critical.
\end{abstract}
\maketitle

\narrowtext

The observation of magnetic quantum critical points and associated
non-Fermi liquid behavior has raised a renewed interest in Kondo
lattice systems~\cite{StewartRMP01}. An antiferromagnetic phase
transition in dimensions higher than one was anticipated very
early on~\cite{Hewson-book93,Doniach77,VarmaNFL02}. For a spin
$S={1 \over 2}$ Kondo lattice system that is not too far away from
half-filling, an antiferromagnetic metal phase is expected if the
RKKY interactions dominate over the Kondo interactions, whereas a
paramagnetic heavy fermion phase should occur in the opposite
limit. Except for some special cases~\cite{TsunetsuguRMP97},
however, there has been a lack of dynamical treatments of the
competition between antiferromagnetism and Kondo effects in
lattice problems. A key issue that has been left open is whether
the zero-temperature magnetic phase transition is first order or
continuous; certainly quantum critical fluctuations can be of any
relevance only if the transition is either continuous or, at most,
very weakly first order. This issue is generally important, but
takes a particularly significant role in the context of the local
quantum critical point (LQCP)~\cite{Lcp-Nature01,GrempelSi02}.
The latter has been invoked to
explain some of the non-Gaussian quantum critical properties
observed in the heavy fermion metals
~\cite{StewartRMP01,Schroder00,Kuchler,Aronson95,Questions01}.

In this paper, we study the dynamical competition between the RKKY
and Kondo interactions using an extended dynamical mean field
theory (EDMFT)~\cite{SiSmith96,Chitra00,Spinglass}.
We analyze the global phase diagram of the model by carrying out
EDMFT studies that cover not only the paramagnetic phase but also
the antiferromagnetically ordered phase. Our analysis involves a
numerical approach to the self-consistent EDMFT equations. To
understand the nature of the quantum phase transition requires a
study at sufficiently low temperatures, and this is in general
very difficult to achieve numerically. We have overcome this
difficulty by focusing on a Kondo lattice model with Ising
anisotropy.
We are able to reach temperatures of the order of about
1\% of the bare Kondo scale, adapting a quantum Monte Carlo (QMC)
method\cite{GrempelRozenberg99,Grempel-QMC98}. Our most important
conclusion is that the zero-temperature transition is indeed
continuous and locally critical. We have also determined
in some detail the dynamics in the ordered and paramagnetic phases.

The
model is specified by the following Hamiltonian:
\begin{equation}
\mathcal{H} = \sum_{ ij\sigma} t_{ij} ~c_{i\sigma}^{\dagger}
c_{j\sigma} + \sum_i J_K ~{\bf S}_{i} \cdot {\bf s}_{c,i}
+ \sum_{ ij} (I_{ij} /2) ~S_{i}^z S_{j}^z .
\label{EQ:kondo-lattice}
\end{equation}
Here, $c_{i\sigma}^{\dagger}$
creates
a conduction electron of spin projection $\sigma$
at the $i$-th site, and ${\bf S}_{i}$ and ${\bf s}_{c,i}$
represent the spins of the $S={1 \over 2}$ local moment and
conduction electrons respectively;
$t_{ij}$ is the hopping integral [corresponding to
a band dispersion $\epsilon_{\bf k}$
]
and $J_K$ the Kondo coupling.
The chemical potential ($\mu$) is chosen so that the system
is less than half-filled;
all the phases we will consider are metallic.
Finally, $I_{ij}$ denote the Ising-exchange interactions between
the local moments. The corresponding Fourier
transform, $I_{\bf q}$, is most negative at an antiferromagnetic
wavevector ${\bf Q}$
($I_{\mathbf{Q}} =-I$).
To access a LQCP, we will consider two-dimensional 
magnetism~\cite{Lcp-Nature01,GrempelSi02} and take
the RKKY -density-of-states in the form:
\begin{equation}
\rho_{I} (\epsilon) \equiv  \sum_{\bf q} \delta ( \epsilon  -
I_{\bf q} ) = (1/{2I})\Theta(I - | \epsilon | ) ,
\label{EQ:rkky-dos}
\end{equation}
where $\Theta$ is the Heaviside function.
In addition, we consider a featureless
$\rho_0(\epsilon)
\equiv  \sum_{\bf k} \delta
( \epsilon  - \epsilon_{\bf k} )$.

We seek to determine the transition into an antiferromagnetic
state.
To
incorporate an
order parameter, we separate the inter-site
exchange 
interaction term into its Hartree and normal-ordered
parts~\cite{SiSmith96,Chitra00}:
$\mathcal{H}_I = \sum_{ ij} I_{ij}
\left [
(1/2)
~:S_{i}^z: : S_{j}^z:
+\langle S_j^z\rangle S_i
-
(1/2) \langle S_i^z\rangle \langle S_j^z \rangle
 \right ]$,
where $:S_{i}^z: \equiv S_{i}^z -\langle S_{i}^z \rangle$. In the
EDMFT mapping to a self-consistent impurity model, the first term
on the RHS of $\mathcal{H}_{I}$
leads to a coupling of
$:S^z:$ (at the selected site) to a fluctuating magnetic field. Up
to constant terms,
the effective impurity action
takes the
form:
\begin{eqnarray}
\mathcal{S}_{\text{imp}}
&=&\mathcal{S}_{\text{top}} + \int_{0}^{\beta} d\tau
[h_{\text{loc}}~ S^{z}(\tau)
+J_K \mathbf{S}(\tau)\cdot \mathbf{s}_{c}(\tau)
] \nonumber \\
&&-\int \int_{0}^{\beta} d\tau d\tau^{\prime}
\sum_{\sigma}
c_{\sigma}^{\dagger}(\tau)G_{0,\sigma}^{-1}
(\tau-\tau^{\prime})c_{\sigma}(\tau^{\prime})
\nonumber \\
&& -\frac{1}{2}\int \int_{0}^{\beta} d\tau
d\tau^{\prime}
S^{z}(\tau)\chi_{0}^{-1}(\tau-\tau^{\prime})S^{z}(\tau^{\prime})\;,
\label{EQ:impurity-action}
\end{eqnarray}
where $\beta=1/k_{B}T$, $\mathcal{S}_{\text{top}}$ describes the
Berry phase of the local moment, and $h_{\text{loc}}$,
$G_{0,\sigma}^{-1}$, and $\chi_{0}^{-1}$ are the static
and
dynamical Weiss fields. These fields are subject to the
self-consistency condition:
\begin{subequations}
\begin{eqnarray}
h_{\text{loc}}&=&-[I-\chi_{0}^{-1}(\omega=0)]~m_{\text{AF}
} \\
\chi_{\text{loc}} (\omega) &=&
\int_{-I}^I d \epsilon ~{\rho_I(\epsilon)
\over {M(\omega)+ \epsilon}},
\label{EQ:self-consistent-b} \\[-1ex]
G_{\text{loc},\sigma} (\omega) &=&
\int_{-D}^{D} d \epsilon ~{\rho_0(\epsilon) \over {\omega + \mu
-\epsilon - \Sigma_{\sigma}(\omega)}} \;,
\end{eqnarray}
\label{EQ:self-consistent}
\end{subequations}
with the aid of Dyson-like equations:
$\Sigma_{\sigma}(\omega)=G_{0,\sigma}^{-1}(\omega) -\frac{1} {G_{
\text{loc},\sigma}(\omega)}$, $ M(\omega)=
\chi_{0}^{-1}(\omega)+\frac{1}{\chi_{\text{loc}}(\omega)}$. Here
$M(\omega)$ and $\Sigma_{\sigma}(\omega)$ are the spin and
conduction-electron self-energies, respectively;
$m_{\text{AF}}=\langle S^{z}\rangle_{\text{imp}}$ is the staggered
magnetization; $\chi_{\text{loc}} (\omega)$ and
$G_{\text{loc},\sigma} (\omega)$ are the Fourier transforms of
$\langle T_{\tau} [:S^z (\tau) ::S^z (0):]\rangle_{\text{imp}}$ and
$-\langle T_{\tau} [c_{\sigma} (\tau) c_{\sigma}^{\dagger}
(0)]\rangle_{\text{imp}}$, respectively.
The lattice spin susceptibility is~\cite{SiSmith96}:
\begin{eqnarray}
\chi({\bf q},\omega) = 1/[I_{\bf q} + M(\omega)]
\label{chi-q-omega}
\end{eqnarray}

The self-consistent impurity model,
Eq.~(\ref{EQ:impurity-action}),
can also be written in a Hamiltonian form,
\begin{eqnarray}
{\cal H}_{\text{imp}} =&& h_{\text{loc}} ~ S^z + J_K ~{\bf S}
\cdot {\bf s}_c + \sum_{p\sigma}
E_{p,\sigma}~c_{p\sigma}^{\dagger}~ c_{p\sigma}
\nonumber\\ &&
+ \; g \sum_{p} S^z \left( \phi_{p} + \phi_{-p}^{\;\dagger}
\right) + \sum_{p} w_{p}\,\phi_{p}^{\;\dagger} {\phi}_{p}\; .
\label{EQ:H-imp}
\end{eqnarray}
It describes
a local moment coupled not only to
a fermionic bath ($c_{p\sigma}$), but also to a dissipative
bosonic bath ($\phi_{p}$)
and a static field.
Here $E_{p\sigma}$, $w_{p}$ and $g$
are such that
$\sum_{p}  { 2 w_{p}g^{2} } / (\omega^{2} - w_{p}^{2})
= - \chi_{0}^{-1}(\omega)$,
$\sum_{p} 1/(\omega - E_{p\sigma})
= G_{0,\sigma}( \omega)$.
In order to understand the quantum phase
transition, we need to reach sufficiently low temperatures;
this is made possible by projecting
Eq.~(\ref{EQ:H-imp}) to a form that does not involve vastly
different energy scales.
For a Kondo lattice, the ordered staggered magnetization
should be  mostly associated with the localized moments.
Its main effect in Eq.~(\ref{EQ:H-imp}) is to generate,
in addition to a finite $h_{\text{loc}}$, a
Zeeman splitting in the fermionic bath.
For featureless conduction electron density of states
we are considering, however, this splitting should
yield a negligible difference between the
fermionic density of states at the chemical potential for
the two spin species.
Denoting $\sum_{p} \delta (\omega - E_{p\sigma}) = N_0$,
we can then adopt the same canonical transformation
${\cal H}_{\text{imp}}'
\equiv U^{\dagger}{\cal H}_{\rm imp} U$
within a bosonization procedure, as used in
Ref.~\cite{GrempelSi02}, to transform
${\cal H}_{\text{imp}}$ into
\begin{eqnarray}
{\cal H}_{\text{imp}}'
 =&&
h_{\text{loc}} ~S^z + \Gamma S^x +
\Gamma_z S^z s^z
+ {\cal H}_0
\nonumber\\ &&
+ \;  g \sum_{p} S^z \left( \phi_{p} + \phi_{-p}^{\;\dagger}
\right) + \sum_{p} w_{p}\,\phi_{p}^{\;\dagger} {\phi}_{p}\;,
\label{EQ:H-imp'}
\end{eqnarray}
where ${\cal H}_0$ describes the fermionic bath, $s^z$ is the
$z$-component of the fermionic bath spin density, and $\Gamma$ and
${\Gamma_z}$ are
respectively determined by the
spin-flip and longitudinal components of the Kondo coupling.
The corresponding partition function is $Z_{\text{imp}}' \sim
{\text{Tr}}~
{\rm exp} [-{\cal S}_{\text{imp}}']$ where
the trace is taken in the spin space and
\begin{eqnarray}
{\cal S}_{\text{imp}}'&=&
\int_0^{\beta} d \tau [
h_{\text{loc}} ~S^z (\tau) + \Gamma S^x
-{1 \over 2}
\int_0^{\beta} d \tau' S^z(\tau) S^z(\tau')
\nonumber\\
&&\times
(\chi_0^{-1}(\tau-\tau') - {\cal K}_c(\tau-\tau') ) ] .
\label{Z-imp}
\end{eqnarray}
Here,
${\cal K}_c (i\omega_n) = \kappa_c |\omega_n| $ ($\kappa_{c}\sim
\Gamma_{z}^{2}N_{0}^{2}$) is a retarded interaction from
integrating out the fermionic bath.

The effective impurity model is solved using a quantum Monte Carlo
(QMC) method~\cite{GrempelRozenberg99,Grempel-QMC98,GrempelSi02}.
We iterate
on $h_{\text{loc}}$ and $\chi_0^{-1}(\tau)$, through the
self-consistency
Eq.~(\ref{EQ:self-consistent}). In the following we will consider
$\kappa_c = \pi$, and $N_0 \Gamma = 0.19$, corresponding to the
bare (normalized) Kondo scale $ N_0 T^0_K = 0.17$. The number of
time
slices in the Trotter decomposition, $N_T$, was chosen
 such as to make the discretization error small enough. We
 found that $N_T \ge 6\;\beta\;T^0_K$ is appropriate.
In our simulations $N_T$ varied between 32 for the higher
temperatures and 512 for the lowest one. We used $10^6$ MC steps
per time slice and performed between 5 and
25
iterations.

\begin{figure}[h]
\centerline{\psfig{file=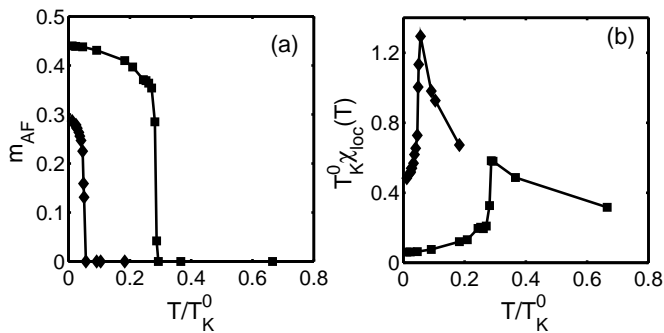,height=4.5cm}}
\caption{%
Temperature dependence of the staggered magnetization $m_
{\text{AF}}$ (a) and the normalized static local susceptibility
(b) for the values of the RKKY interaction  $I/T_{K}^{0}=3.0$
(squared curves)
and $1.5$
(diamond curves).
  }
\label{FIG:MafT-ChiT}
\end{figure}

That antiferromagnetism develops for sufficiently large RKKY
interactions is seen in Fig.~\ref{FIG:MafT-ChiT}(a), which shows
the temperature dependence of the staggered
magnetization
($m_{\text{AF}}$) as a function of temperature at $I=3 T_K^0$ and
$I=1.5 T_K^0$. The order parameter $m_{\text{AF}}$ becomes
non-zero at $T_{ \text{N}} \approx 0.29 T_K^0$ and $T_{\text{N}}
\approx 0.05T_K^0$, respectively. The numerical data are
consistent with a jump in $m_{{\text{AF}}}$ at $T_{\text{N}}$,
suggesting that the finite temperature
 magnetic transition is first order. This
is further supported by the $T$-dependence of the static local
susceptibility shown in Fig.~\ref{FIG:MafT-ChiT}(b). It is seen
that $\chi_{\text{loc}}(T_{\text{N}})$
is finite whereas,
for an RKKY density of states of Eq.~(\ref{EQ:rkky-dos}),
it must be divergent
if the transition were second order
[following arguments similar to those given after
Eq.~(\ref{EQ:qcp-1})]~\cite{note2}.

\begin{figure}[t]
\centerline{\psfig{file=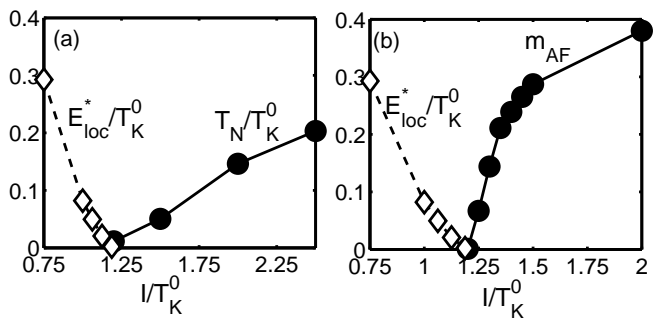,width=9cm}}
\caption{%
(a) The N\'eel temperature (solid dots) and $E_{\text{loc}}^*$
(open diamonds) as a function of $I/T_K^0$, the ratio of the RKKY
interaction to the bare Kondo scale; (b) The staggered
magnetization (solid dots) as a function of of $I/T_K^0$, at
$T=0.011T_K^0$. $E_{\text{loc}}^*$ characterizes the coherence
scale of the paramagnetic solution (see the main text). The solid
and dashed lines are guides to eyes. } \label{FIG:Eloc-Tn}
\end{figure}

The dependence of the N\'eel temperature on the RKKY interaction
is shown in Fig.~\ref{FIG:Eloc-Tn}(a). It is seen that
$T_{\text{N}}$ vanishes at $I_c /T_K^0 \approx 1.2$. The staggered
magnetization at the lowest temperature ($T=0.011T_K^0$) also
vanishes continuously at the same value $I_c$ as shown in
Fig.~\ref{FIG:Eloc-Tn}(b). Finally, the size of the jump of
$m_{\text{AF}}$ at $T_{\text{N}}$ decreases with the latter as the
strength of the RKKY interaction is reduced. The extrapolated
value of the jump also goes to zero at $I_c$. These observations
strongly suggest that the zero-temperature transition is second
order.


We also show in
Fig.~\ref{FIG:Eloc-Tn}
$E_{\text{loc}}^*$, the coherence scale of the Kondo lattice as a
function of $I$. It may be seen that
 $E_{\text{loc}}^*$
is finite for $I<I_c$ but decreases with increasing $I$;
within our numerical uncertainty, it vanishes
precisely at $I_c$.
$E_{\text{loc}}^*$ characterizes the crossover from the low energy
Fermi liquid regime, in which the local susceptibility has a Pauli
form, to the high energy quantum critical regime where the local
susceptibility is logarithmically singular.
This scale is determined numerically from the frequency
dependence of the local susceptibility
as discussed in Ref.~\cite{GrempelSi02}.
In this range  ($I<I_c$), the solution derived from starting
the iteration with a zero trial $h_{\text{loc}}$ is
identical to that with a finite trial $h_{\text{loc}}$.
That the values of $I_c$ determined from the paramagnetic and ordered
sides are the same provides an additional evidence for the second order
nature of the quantum phase transition.

\begin{figure}[t]
\centerline{\epsfxsize=75mm\epsfbox{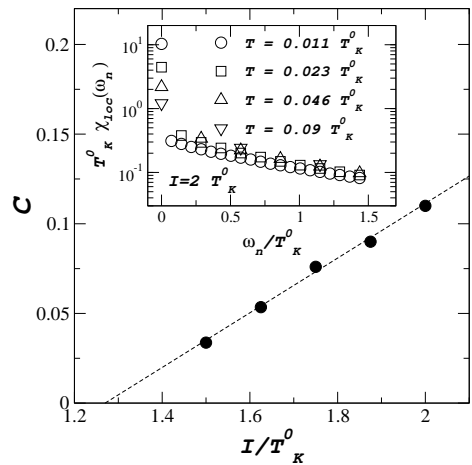}}
\caption{%
Curie constant ${\cal C}$
in the nominally ``paramagnetic'' solution
as a function of $I/T_K^0$. Inset: the local susceptibility
vs. Matsubara frequency $\omega_n$ for $I=2T_K^0$ at various
temperatures; the jump at $\omega_n=0$ measures ${\cal C}/T$.
}
\label{FIG:Curie}
\end{figure}

We wish to stress a subtle issue in the determination of $I_c$
coming from the paramagnetic side. If we apply the paramagnetic
EDMFT equations (i.e., with
$h_{\text{loc}}=0$) to $I>I_c$, we find that
they would
converge to a nominally ``paramagnetic'' solution
(at least for $I\le 3 T^0_K$).
These solutions are, however, characterized by
free residual (but not ordered!) moments,
as signaled by
a ${\cal C}/T$ divergence
in
$\chi_{\rm loc}(T)$
at low temperatures;
the constant ${\cal C}$ is an effective Curie constant.
This
Curie constant
manifests
also in the behavior of $\chi_{\text{loc}}(\omega_n)$
at $\omega_n=0$, as illustrated (for a particular
value $I=2T_K^0 \gg I_c$) in the inset to
Fig.~\ref{FIG:Curie}:
$\chi_{\text{loc}}(\omega_n=0)$ jumps by an amount ${\cal C}/T$
compared to the value extrapolated from $\chi_{\text{loc}}(\omega_n)$
at finite $\omega_n$~\cite{note}.
(Without going to low temperatures, this
jump would have been hard to see.)
From the dependence of ${\cal C}$ on $I$ (Fig.~\ref{FIG:Curie}, main plot),
the Curie constant is seen to vanish as $I$ approaches $I_c$ from
above.
The finite Curie constant for all $I>I_c$
shows that, in this region, the nominally ``paramagnetic'' solution
is unstable and the true ground state
is
the antiferromagnetic solution.

That $E_{\text{loc}}^*$ has to vanish at $I_c$ can in fact be
understood once the continuous onset of $T_{\text{N}}$ and
$m_{\text{AF}}$ at $I_c$ is established.
A second order magnetic transition at zero temperature
means that $\chi({\bf Q},\omega=0,T=0)$ diverges at $I_c$.
Using
\begin{eqnarray}
\chi({\bf Q},\omega=0,T=0) = 1/[M(\omega=0,T=0)-I]
\label{EQ:qcp-1}
\end{eqnarray}
[which can be seen by inserting $I_{\bf Q} = -I$ into
Eq.~(\ref{chi-q-omega})],
this implies that $M(\omega=0,T=0)=I$.
It then follows from the
self-consistency Eq.~(\ref{EQ:self-consistent-b}), together with
Eq.~(\ref{EQ:rkky-dos}), that the local susceptibility
$\chi_{\text{loc}}(\omega=0,T=0)$ also diverges~\cite{note2} at
$I=I_c$. This places the local Kondo problem, as specified by
Eq.~(\ref{EQ:H-imp}) with $h_{\text{loc}}=0$, to be at its own
critical point~\cite{note2}. In other words, the criticality of
the Kondo physics is embedded in the criticality of the magnetic
transition in the lattice; the corresponding coherence energy
scale, $E_{\text{loc}}^*$, vanishes.

The overall phase diagram is now specified by
Fig.~\ref{FIG:Eloc-Tn}(a). While the first order nature of the
finite temperature transition is an artifact of EDMFT, reflecting
the fact that EDMFT contains no spatial anomalous dimension, the
LQCP at zero temperature is expected to be
robust~\cite{Lcp-Nature01,note2}. By establishing the continuous
nature of the zero-temperature transition, we can identify a
finite-temperature quantum-critical ``fan'' that is anchored by
the ($T=0$) LQCP at $I=I_c$, even in the absence of a correct
understanding of the finite temperature transition. This quantum
critical fan can be used to describe the non-Gaussian quantum
critical behavior observed in heavy fermion metals.

\begin{figure}[t]
\centerline{\psfig{file=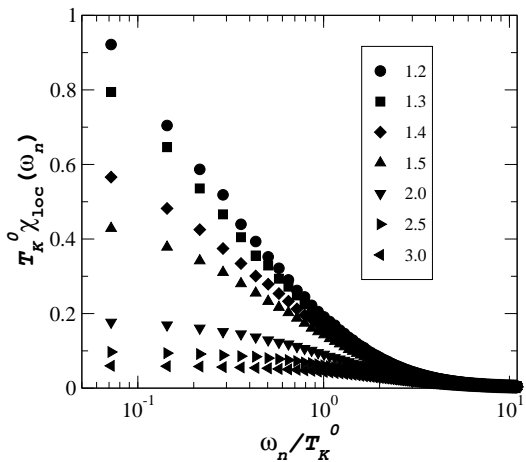,width=7cm}}
\caption{%
The normalized local susceptibility vs. Matsubara frequency,
at $T=0.011T_K^0$.
The legend specifies $I/T_K^0$.
} \label{FIG:Chi_omega}
\end{figure}

To better understand the dynamics in the ordered phase, we have
also calculated the frequency dependence of the local
susceptibility. Fig.~\ref{FIG:Chi_omega} plots
$\chi_{\text{loc}}(\omega_n)$ vs $\omega_n$ for various values of
$I/T_K^0$, at $T=0.011T_K^0$. On the ordered side, the peak
susceptibility, $\chi({\bf Q},\omega_n)$, saturates to a finite
value as a result of the finite magnetic order parameter.
Correspondingly, the local susceptibility,
$\chi_{\text{loc}}(\omega_n)$, which is equal to the ${\bf
q}$-average  of
$\chi({\bf q},\omega_n)$, also saturates.
The saturated value of
$\chi_{\text{loc}}(\omega_n)$ increases as the RKKY interaction
and, correspondingly, $h_{\text{loc}}$, is reduced. At
$I=I_c\approx 1.2 T_K^0$, $h_{\text{loc}}$ goes to zero and the
local susceptibility becomes logarithmically singular, the form
that is associated with the critical point of this self-consistent
impurity model; the result
at $I=I_c$ agrees with that of the
analysis from the paramagnetic side.


One of us (JXZ)
acknowledges the useful discussions with A. V. Balatsky, Y. Bang
and L. Zhu,
and
the hospitality of Rice University, where part of the research was
carried out. This work has been supported by US DOE
(JXZ),
NSF at KITP (DRG)
and NSF,
the Welch Foundation and TCSAM (QS). It was first
reported in Ref.~\cite{APS03}. We have recently learned that P.
Sun and G. Kotliar~\cite{Sun03} have independently
addressed similar issues in an Anderson lattice at high temperatures.


\begin{thebibliography}{99}
\bibitem{note2} Both a finite
$\chi_{\text{loc}}(T_N \ne 0, I>I_c)$
and a divergent
$\chi_{\text{loc}}(T = 0, I=I_c)$
are compatible with the fact that
$\chi_{\text{loc}}$, being also determined by
an effective impurity model, must be finite at
any finite $T$ but can diverge at $T=0$.
In general, for a ($T=0$) LQCP a vanishing spatial anomalous
dimension $\eta$ is internally consistent~\cite{Lcp-Nature01}
so our results should be robust beyond EDMFT.
A classical critical point, on the other hand, usually
has a finite $\eta$ and is not correctly described by EDMFT.
\bibitem{note} The solution with $h_{\text{loc}}=0$ and
a finite Curie constant at $I>I_c$ can be understood as follows. A
${ {\cal C} \over T } \delta_{n,0}$ term in
$\chi_{\text{loc}}(\omega_n)$ implies an extra term in the Weiss
field of the form $a(T) \delta_{n,0}$, where $a(T)$ is
positive.
The solution for
$\chi_{\text{loc}}(\tau)$ to such an Ising Bose-Fermi Kondo
model~\cite{Zhu02} contains connected and disconnected
pieces; self-consistently, the two pieces determine the Curie
constant and regular part of $\chi_{\text{loc}}(\omega_n)$,
respectively.
\end{thebibliography}
\end{document}